%% file: main.tex
\newcommand{\CLASSINPUTtoptextmargin}{19mm}%
\newcommand{\CLASSINPUTbottomtextmargin}{43mm}%
\newcommand{\CLASSINPUTinnersidemargin}{12.9mm}%
\newcommand{\CLASSINPUToutersidemargin}{12.9mm}%

\documentclass[conference,10pt,a4paper]{IEEEtran}
\usepackage{amsmath}
\usepackage{xcolor}
\usepackage{times}
\usepackage{graphicx}
\usepackage{multirow}
\usepackage[none]{hyphenat}
\usepackage{float}
\usepackage{iftex}
\usepackage{algorithm}
\usepackage{algpseudocode}
\usepackage{amsmath}
\usepackage{mathtools} 
\usepackage{psfrag}
\usepackage{subcaption}
\usepackage[mode=match,per-mode=symbol]{siunitx}
\usepackage{hhline}
\usepackage{trfsigns}
\usepackage{amsfonts}
\usepackage{acronym}
\usepackage{nohyperref}  

\newacro{3GPP}{3rd Generation Partnership Project}
\newacro{5G}{fifth generation}
\newacro{5G NR}{Fifth Generation New Radio}
\newacro{6G}{sixth generation}
\newacro{A/D}{analog-to-digital}
\newacro{AAL}{array aperture line}
\newacro{ABE}{analog back-end}
\newacro{ADC}{analog-to-digital converter}
\newacro{AFE}{analog front-end}
\newacro{AGV}{automatic guided vehicle}
\newacro{AM-AM}{amplitude-to-amplitude modulation}
\newacro{AM-PM}{amplitude-to-phase modulation}
\newacro{AWGN}{additive white Gaussian noise}
\newacro{B5G}{beyond \ac{5G}}
\newacro{BB}{baseband}
\newacro{BER}{bit error ratio}
\newacro{BPSK}{binary phase-shift keying}
\newacro{BP}{band-pass}
\newacro{BS}{base station}
\newacro{CDM}{code-division multiplexing}
\newacro{CFO}{carrier frequency offset}
\newacro{CFR}{channel frequency response}
\newacro{CIR}{channel impulse response}
\newacro{CoMP}{coordinated multipoint}
\newacro{CP}{cyclic prefix}
\newacro{CPE}{common phase error}
\newacro{CPO}{carrier phase offset}
\newacro{CRLB}{Cram\'er-Rao lower bound}
\newacro{CS}{chirp sequence}
\newacro{CSI}{channel state information}
\newacro{CW}{continuous wave}
\newacro{CZT}{chirp Z-transform}
\newacro{D/A}{digital-to-analog}
\newacro{DAC}{digital-to-analog converter}
\newacro{DDC}{digital down-conversion}
\newacro{DDS}{direct digital synthesis}
\newacro{DFRC}{dual-function radar-communication or dual-functional radar-communication}
\newacro{DFnT}{discrete Fresnel transform}
\newacro{DFT}{discrete Fourier transform}
\newacro{DL}{downlink}
\newacro{DMRS}{demodulation reference signal}
\newacro{DoA}{direction-of-arrival}
\newacro{DoD}{direction-of-departure}
\newacro{DPD}{digital pre-distortion}
\newacro{DUC}{digital up-conversion}
\newacro{ETSI}{European Telecommunications Standards Institute}
\newacro{EVM}{error vector magnitude}
\newacro{FDE}{frequency-domain equalization}
\newacro{FDM}{frequency-division multiplexing}
\newacro{FO}{frequency offset}
\newacro{FR2}{Frequency Range 2}
\newacro{gNB}{gNodeB}
\newacro{HP}{high-pass}
\newacro{IBFD}{in-band full duplex}
\newacro{ICI}{intercarrier interference}
\newacro{IDFT}{inverse discrete Fourier transform}
\newacro{IDFnT}{inverse discrete Fresnel transform}
\newacro{IF}{intermediate frequency}
\newacro{IHE}{Institute of Radio Frequency Engineering and Electronics}
\newacro{I/Q}{in-phase/quadrature}
\newacro{IBO}{input back-off}
\newacro{IM3}{third-order intermodulation}
\newacro{IP1dB}{input-referred 1-dB compression point}
\newacro{ISAC}{integrated sensing and communication}
\newacro{ISI}{intersymbol interference}
\newacro{ISLR}{integrated-sidelobe level ratio}
\newacro{IoT}{Internet of Things}
\newacro{JCAS}{joint communication and sensing}
\newacro{KIT}{Karlsruhe Institute of Technology}
\newacro{KPI}{key performance indicator}
\newacro{LDPC}{low-density parity-check}
\newacro{LFSR}{linear-feedback shift register}
\newacro{LNA}{low-noise amplifier}
\newacro{LO}{local oscillator}
\newacro{LoS}{line-of-sight}
\newacro{LP}{low-pass}
\newacro{LPI}{low probability of intercept}
\newacro{LS}{least squares}
\newacro{mmWave}{milimeter wave}
\newacro{MIMO}{multiple-input multiple-output}
\newacro{MLE}{maximum likelihood estimator}
\newacro{MLS}{maximum-length sequence}
\newacro{MRC}{maximal-ratio combining}
\newacro{MUSIC}{multiple signal classification}
\newacro{NAF}{normalized angular frequency}
\newacro{NB}{narrowband}
\newacro{NLoS}{non-line-of-sight}
\newacro{NR}{new radio}
\newacro{OCDM}{orthogonal chirp-division multiplexing}
\newacro{OFDM}{orthogonal frequency-division multiplexing}
\newacro{OOB}{out-of-band}
\newacro{OTA}{over-the-air}
\newacro{P/S}{parallel-to-serial}
\newacro{PA}{power amplifier}
\newacro{PACF}{periodic autocorrelation function}
\newacro{PAPR}{peak-to-average power ratio}
\newacro{PCCF}{periodic cross-correlation function}
\newacro{PCL}{passive coherent location}
\newacro{PLC}{powerline communication}
\newacro{PLL}{phase-locked loop}
\newacro{PMCW}{phase-modulated continuous wave}
\newacro{PMN}{perceptive mobile network}
\newacro{PN}{oscillator phase noise}
\newacro{PoC}{proof-of-concept}
\newacro{PPLR}{peak power loss ratio}
\newacro{PRBS}{pseudorandom binary sequence}
\newacro{PRS}{positioning reference signal}
\newacro{PSD}{power spectral density}
\newacro{PSF}{point spread function}
\newacro{PSLR}{peak-to-sidelobe level ratio}
\newacro{QPSK}{quadrature phase-shift keying}
\newacro{RadCom}{radar-communication}
\newacro{RCS}{radar cross section}
\newacro{RF}{radio-frequency}
\newacro{RFS}{random finite set}
\newacro{RIS}{reflective intelligent surface}
\newacro{RMS}{root mean square}
\newacro{RMSE}{root mean squared error}
\newacro{RX}{receiver}
\newacro{SC}[S\&C]{Schmidl \& Cox}
\newacro{SFO}{sampling frequency offset}
\newacro{SIC}{self-interference cancellation}
\newacro{SINR}{signal-to-interference-plus-noise ratio}
\newacro{SIR}{signal-to-interference ratio}
\newacro{SISO}{single-input single-output}
\newacro{SJ}{sampling jitter}
\newacro{SNR}{signal-to-noise ratio}
\newacro{SoC}{system-on-a-chip}
\newacro{SQNR}{signal-to-quantization-noise ratio}
\newacro{SSB}{synchronization signal block}
\newacro{STO}{symbol time offset}
\newacro{S/P}{serial-to-parallel}
\newacro{TDD}{time-division duplexing}
\newacro{TDE}{time-domain equalization}
\newacro{TDM}{time-division multiplexing}
\newacro{TDR}{time-domain reflectometry}
\newacro{TITO}{tilt inference of time offset}
\newacro{TO}{time offset}
\newacro{TR}{technical report}
\newacro{TS}{technical specification}
\newacro{TX}{transmitter}
\newacro{UE}{user equipment}
\newacro{UL}{uplink}
\newacro{ULA}{uniform linear array}
\newacro{V2V}{vehicle-to-vehicle}
\newacro{ZF}{zero forcing}
\newacro{ZP}{zero padding}
\ifPDFTeX%
\usepackage{t1enc}
\usepackage{times}
\fi%
\ifXeTeX%
\usepackage{fontspec}
\fi%
\ifLuaTeX%
\usepackage{fontspec}
\fi%
\input{EuMW_modify_IEEEtran_18b_CTAN_V5}

\usepackage{fancyhdr}
\fancypagestyle{empty}{fancy}

\begin{document}
\raggedbottom
%
%
%
\title{Secure OFDM Waveform Design for ISAC: Artificial Phase-Doppler Shifts Against Passive Sensing}
%
%
\author{%
\IEEEauthorblockN{%
Umut Utku Erdem\IEEEauthorrefmark{1}, 
Lucas Giroto\IEEEauthorrefmark{1}\IEEEauthorrefmark{2}, 
Tobias Chaloun\IEEEauthorrefmark{3}, 
Tom Schipper\IEEEauthorrefmark{3},\\
Taewon Jeong\IEEEauthorrefmark{1},
Christian Karle\IEEEauthorrefmark{4},
Benjamin Nuss\IEEEauthorrefmark{5},
Thomas Zwick\IEEEauthorrefmark{1}
}
\IEEEauthorblockA{%
\IEEEauthorrefmark{1}Institute of Radio Frequency Engineering and Electronics, Karlsruhe Institute of Technology, Germany\\
\IEEEauthorrefmark{2}Nokia Bell Labs Stuttgart, Germany\\
\IEEEauthorrefmark{3}Hensoldt Sensors GmbH, Germany\\
\IEEEauthorrefmark{4}Institute for Information Processing Technology, Karlsruhe Institute of Technology, Germany\\
\IEEEauthorrefmark{5}Professorship of Microwave Sensors and Sensor Systems, Technical University of Munich, Germany\\
E-mail: umut.erdem@kit.edu
}
}
%
\maketitle
%
%
%
\begin{abstract}
This paper proposes a novel low probability of intercept (LPI) waveform design approach for orthogonal frequency-division multiplexing (OFDM)-based integrated sensing and communication systems by introducing artificial phase and Doppler shifts. These controlled impairments, unknown to eavesdroppers, effectively disrupt passive radar processing and intercept attempts. At legitimate receivers, they can be fully compensated, so that standard OFDM communication and sensing performance are preserved. To support the effectiveness of the proposed LPI waveform design for OFDM-based ISAC, measurement results with 1\,GHz bandwidth at 27\,GHz are presented considering different impairment introduction approaches, all with no impact on cooperative system performance, and compensation capabilities at the eavesdropper.
\end{abstract}
%
%

\section{Introduction}

\acused{ISAC}Integrated sensing and communication (ISAC) systems are emerging as a key technology for \ac{6G} networks, enabling simultaneous high-data-rate communication and accurate sensing over the same infrastructure. 
However, \ac{ISAC} systems are exposed to security challenges as wireless may be intercepted by eavesdroppers \cite{PLS_survey}. 
Unlike communication-only or radar-only systems, \ac{ISAC} presents security challenges that can be classified into communication security and sensing security categories. Communication security arises from the possibility that embedded information within the \ac{ISAC} waveform may be exposed to malicious nodes, which could function as eavesdroppers. Additionally, eavesdroppers could exploit the transmitted \ac{ISAC} waveform reflected by targets to extract information on the environment via passive sensing \cite{Adam_Ksi_2023}. These security issues must be addressed during waveform design.

Among possible waveforms for \ac{ISAC}, \ac{OFDM} is a strong candidate due to its spectral efficiency, flexibility, and compatibility with existing communication standards such as, \ac{5G NR}, and IEEE 802.11ad \cite{Zu_ISAC_challenges}.
Yet, its standardized, predictable structure (e.g., pilots, preambles) lacks inherent robustness against passive sensing attacks. While some passive sensing attacks rely on cross-correlating the surveillance signal with a clean reference signal (\ac{PCL}) \cite{Adam_Ksi_2023}, the explicit pilot structure of the waveform allows adversaries to bypass this complex requirement. By simply exploiting known pilot locations, an eavesdropper can synchronize and estimate the channel response using standard receiver algorithms. This structure-based vulnerability necessitates the design of specific \ac{LPI} features.

Several \ac{LPI} strategies for \ac{OFDM} have been investigated previously. One approach involves subcarrier power allocation to minimize transmit power and detection probability while maintaining radar performance. This method relies on prior scene knowledge, making it mainly suitable for tracking. For example, \cite{shi-1} optimizes power for \ac{ISAC} to analyze range estimation accuracy. However, for search mode with unknown targets, this is unsuitable. Alternatively, symbol-level variations, such as different \ac{CP} lengths \cite{Bouanen}, can degrade eavesdropper synchronization. While offering security, these methods often degrade legitimate receiver performance or increase computational complexity \cite{Zhang}.

In this paper, an \ac{LPI} waveform for \ac{OFDM}-based \ac{ISAC} is proposed using artificial phase and Doppler shifts to degrade both communication demodulation and radar processing at unauthorized receivers. The proposed algorithm is particularly effective against undesired passive bistatic radar sensing, as eavesdroppers cannot fully compensate for unknown artificial parameters despite having access to the transmitted waveform structure up to some extent. Measurement results demonstrate the effectiveness of the proposed method, showing higher \ac{ISLR} and \ac{PSLR}, and lower \ac{SINR} at the eavesdropper, while maintaining ideal performance at the legitimate receiver.


\section{System Model}
\label{sec:system-model}

\begin{figure}[t]
\centering
\includegraphics[width=0.9\columnwidth]{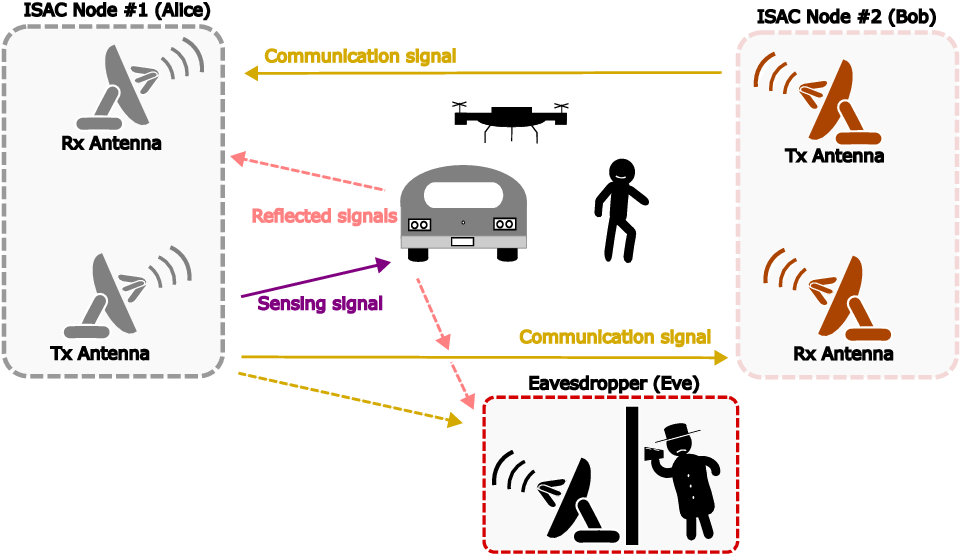}
\caption{Multi-node ISAC system model where an eavesdropper (Eve) performs passive sensing using the signals transmitted from ISAC Node \#1 (Alice)}
\label{fig:system}
\vspace{-0.2cm}
\end{figure}

In this section, a multi-node \ac{ISAC} system with multiple transceiver nodes that simultaneously transmit data and perform sensing is considered. As illustrated in Fig.~\ref{fig:system}, \ac{ISAC} Node  \#1 and  \ac{ISAC} Node \#2, denoted as Alice and Bob respectively, transmit \ac{OFDM} signals for both
communication and radar purposes. The environment has
\mbox{$P \in \mathbb{N}_{>0}$} radar targets (e.g., vehicles, pedestrians, drones), off which the aforementioned signals reflect. In addition to the cooperative \ac{ISAC} nodes,
an eavesdropper, denoted as Eve, is present in
the environment. Eve receives both the direct \ac{LoS}
communication signal from Alice and the radar echoes from targets. Eve operates as a passive bistatic radar with respect to Alice and utilizes standard OFDM channel estimation processing to perform bistatic sensing, rather than employing a dedicated and complex \ac{PCL} processing chain \cite{Adam_Ksi_2023}. 

Let \mbox{$X[k,m] \in \mathbb{C}$} denote the complex modulation symbols transmitted on subcarrier \mbox{$k \in \{0, \dots, N{-}1\}$} and \ac{OFDM} symbol \mbox{$m \in \{1, \dots, M\}$} forming the frequency-domain transmit frame \mbox{$\mathbf{X} \in \mathbb{C}^{N \times M}$}. The subcarrier spacing is \mbox{$\Delta f = B/N$}, and the baseband sampling interval is \mbox{$T_\text{s} = 1/B$}, where $B$ denotes the bandwidth. The frame $\mathbf{X}$ contains pilot subcarriers used for channel estimation and synchronization. Regular pilot lattice is assumed where pilots are spaced by \mbox{$\Delta N_{\text{pil}} \in \mathbb{N}_{>0}$} subcarriers in frequency and \mbox{$\Delta M_{\text{pil}} \in \mathbb{N}_{>0}$} symbols in time. Consequently, the total number of pilot-bearing \ac{OFDM} symbols $M_{\text{pil}}$ and subcarriers $N_{\text{pil}}$ in the frame are \cite{giroto2024_tmtt} \begin{equation}
    M_{\text{pil}} = \left\lfloor {M}/{\Delta M_{\text{pil}}} \right\rfloor + 1, \quad N_{\text{pil}} = \left\lfloor {N}/{\Delta N_{\text{pil}}} \right\rfloor + 1,
\end{equation}
where $\lfloor\cdot\rfloor$ denotes the floor operation. In this paper, it is assumed that Eve knows the transmitted complex signals at pilot-bearing \ac{OFDM} symbols and subcarriers and use those for radar processing.

To enhance the \ac{LPI} properties of the transmitted signal, Alice can apply phase rotations to the complex modulation symbols. Let $\mathbf{\Phi} \in \mathbb{R}^{N \times M}$ be the artificial phase matrix shared between Alice and Bob, where each element of $\mathbf{\Phi}$  is denoted as $\phi_{k,m}$. Each complex modulation symbol $X[k,m]$ is rotated by a random phase shift $\phi_{k,m} \sim \mathcal{U}(-\pi, \pi)$, where $\mathcal{U}(a, b)$ denotes uniform sampling from interval $(a,b)$, as follows
    \begin{equation}
    X_\text{L}[k,m] = X[k,m] \cdot \e^{\im \phi_{k,m}}.
    \label{eq:elementwise_rotation}
    \end{equation}
Next, the \ac{IDFT} is applied on $X_\text{L}[k,m]$ and the time-domain signal for the $m$th \ac{OFDM} symbol (\ac{CP}-free) is given by, 
\begin{equation}
x^m[n] = \frac{1}{\sqrt{N}} \sum_{k=0}^{N-1} X_\text{L}[k,m] \e^{\im 2 \pi kn/N}, \quad n = 0, \dots, N{-}1.
\end{equation}
A \ac{CP} of length $N_{\text{cp}}$ is appended to $x^m[n]$ to prevent \ac{ISI}. The resulting time-domain signal is denoted as $s^m[n]$.
 In addition to introduced phase rotations $\phi_{k,m}$ in frequency domain, Alice can also introduce artificial Doppler shifts based on the virtual Doppler shift vector $\mathbf{f}_{\text{D},\text{v}} \in \mathbb{R}^{\text{M}}$. Let $f^m_{\text{D},\text{v}} \sim \mathcal{U}(-\Delta f/2,\Delta f/2)$ be the $m$th entry in $\mathbf{f}_{\text{D},\text{v}}$. Consequently, the time domain signal (including \ac{CP}) after per-symbol Doppler shift is 
\begin{equation}
    s^{m}_\text{A}[n] = s^{m}[n] \e^{\im 2\pi f_{\text{D},\text{v}}^{m} n T_\text{s}}, \quad n = 0, \dots, N+N_{\text{cp}}-1.
    \label{eq:time_Doppler}
\end{equation}
The resulting discrete-time domain \ac{OFDM} symbols $s^{m}_\text{A}[n]$ first undergo parallel-to-serial conversion, followed \ac{D/A} conversion, and upconversion to the desired \ac{RF} carrier. The resulting signal then propagates through the physical environment, where it is attenuated, reflected off $P$ radar targets associated with different ranges and Doppler shifts due to their motion. 

In this paper, \ac{OFDM}-symbol-wise phase and Doppler impairments are also considered as baseline to evaluate the impact of different impairment strategies. This simplified \ac{LPI} strategy is also considered where both the artificial phase and Doppler distortions are applied at the symbol level:
\begin{equation}
X_\text{L}[k,m] = X[k,m] \cdot \e^{\im\phi_m}, \quad s^m_A[n] = x_m[n] \cdot \e^{\im2 \pi f^m_{\text{D},\text{v}} n T_s}.
\label{eq:symbol_wise}
\end{equation}
Although this scheme offers limited \ac{LPI} strength compared to subcarrier-wise phase rotation design in \eqref{eq:elementwise_rotation}, it serves as a baseline for analysis. Measurement results in Section~\ref{sec:results} explore both strategies.
\subsection{Legitimate Receiver Processing}
At the legitimate receiver, either Alice or Bob, the signal is first bandpass filtered, amplified by \ac{LNA} and downconverted to baseband, finally undergoing \ac{A/D} conversion. The discrete-time equivalent model for the received echo signal during $m$th symbol interval $y^m[n]$ can be expressed as,
\begin{equation}
y^m[n] \approx \sum_{p=1}^{P} \alpha_p \, s^m_\text{A}[n - n_p] \cdot \e^{\im2\pi f_{\mathrm{D},p}(m(N+N_{\text{cp}})+n)T_\text{s}} + w^m[n],
\label{eq:rx_total}
\end{equation}
where $\alpha_p$ and $f_{\text{D},p}$ are the complex reflection coefficient and Doppler shift associated with the $p$th target, \mbox{$n_p = \lfloor \tau_p / T_s \rceil$} is the delay in samples, where $\lfloor\cdot\rceil$ is the rounding operation, and $w^m[n]$ is additive white Gaussian noise (AWGN) during the $m$th symbol interval.

The legitimate receiver, possessing \mbox{$\{\mathbf{\Phi}, \mathbf{f}_{\text{D},\text{v}}^{m}\}$}, first removes the artificial Doppler shift $f_{\text{D},\text{v}}^{m}$ from the received signal $y^m[n]$ by performing
\begin{equation}
    \tilde{y}^m[n] = y^m[n] \e^{-\im 2\pi f_{\text{D},\text{v}}^{m} n T_\text{s}}.
    \label{eq:dopp_comp}
\end{equation}
 After serial-to-paralel conversion, and removing the \ac{CP} at each \ac{OFDM} symbol, an $N$-point \ac{DFT} can be performed to obtain \mbox{$\tilde{Y} \in \mathbb{C}^{N\times M}$}, which indicates the received frequency domain frame before phase compensation.
 The legitimate receiver removes the artificial phase shifts $\phi_{k,m}$ at each entry of $\tilde{Y}$, denoted as $\tilde{Y}[k,m]$, as follows
\begin{equation}
    Y_\text{comp}[k,m] = \tilde{Y}[k,m] \e^{-\im \phi_{k,m}}.
\end{equation}
After phase rotation compensation, the legitimate receiver computes the channel response \mbox{$D[k,m] = Y_\text{comp}[k,m]/X[k,m]$} via frequency-domain equalization.
However, after \eqref{eq:dopp_comp}, the residual phase term $\e^{-\im2\pi f_{\text{D},\text{v}}^{m}n_pT_\text{s}}$ occurs at the $m$th \ac{OFDM} symbol for target $p$. Since the artificial Doppler shift is compensated based on instant $[n]$, the residual phase rotation appears as a function of the delay for each target. This residual phase term, which changes at every symbol, should be corrected during the radar processing which can lead to phase jumps among \ac{OFDM} symbols during Doppler processing. 
Since the residual phase term is a function of delay of each target, it can be corrected in the range profile $I_\text{r}[r,m] \in \mathbb{C}^{N\times M}$, which is the \ac{IDFT} of $D[k,m]$ across the subcarriers. As the each range bin $r \in \{0,\dots,N-1\}$ in $I_\text{r}[r,m]$ corresponds to delay $\tau _r = r/B$, the following range-bin dependent correction
\begin{equation}
    \hat{I}_r[r,m] = I_\text{r}[r,m] \e^{\im 2\pi f_{\text{D},\text{v}}^{m} \tau_r}.
    \label{eq:range_correction}
\end{equation}
can be applied to remove the residual phase term caused by delay.
 Finally, a \ac{DFT} is applied to $\hat{I}_\text{r}[r,m]$ across the \ac{OFDM} symbols to generate the range-Doppler map.

\subsection{Eavesdropper Processing}
Eve is modeled as a passive bistatic receiver with perfect synchronization to the transmitter, and two distinct processing strategies are considered for Eve: 
\begin{itemize}
    \item \textit{Blind Eve:} Operating without prior knowledge of the specific LPI strategy, Eve assumes a standard \ac{OFDM} waveform and attempts to process the signal directly by computing the channel response by \mbox{$D_\text{pil}[k,m] = R_\text{pil}[k,m]/X_\text{pil}[k,m]$} for pilot subcarriers only. 
    When \ac{OFDM}-symbol-wise phase rotation and Doppler shifts were introduced at the transmitter, this processing strategy will lead to an approximate processing gain loss of $M_\text{pil}$ and also lead to strong distortions along the Doppler axis. When subcarrier-wise phase rotation and \ac{OFDM}-symbol-wise Doppler shift strategy is adopted at the transmitter, this processing will lead to a completely distorted radar image.

    \item \textit{Smart Eve:} Possessing prior knowledge of the LPI framework, Eve attempts to estimate the artificial parameters. Specifically, Eve computes \mbox{$I_\text{r}^\text{eve}\in \mathbb{C}^{N_\text{pil}\times M_\text{pil}}$} by calculating the \ac{IDFT} of $D_\text{pil}[k,m]$ along the subcarriers. Eve attempts to track and remove the random phase shifts by aligning the phase of the strong \ac{LoS} path signal, which appears at $r=0$ in the range profile assuming perfect synchronization w.r.t. to it \cite{giroto2024_tmtt}, across \ac{OFDM} symbols. This processing strategy can compensate artificial phase impairments when \ac{OFDM}-symbol-wise phase and Doppler shifts are introduced by the transmitter as described in \eqref{eq:symbol_wise}. On the other hand, when subcarrier-wise phase rotations are introduced, Eve's compensation strategy would lead to a highly distorted radar image since $\phi_{k,m}$ changes for each pilot, and the estimated phase for the \ac{LoS} path cannot be translated into the pseudorandom phases for each pilot subcarrier.
    
\end{itemize}

\section{Measurement Results}

\begin{figure}[t]
\centering
\includegraphics[width=1\columnwidth]{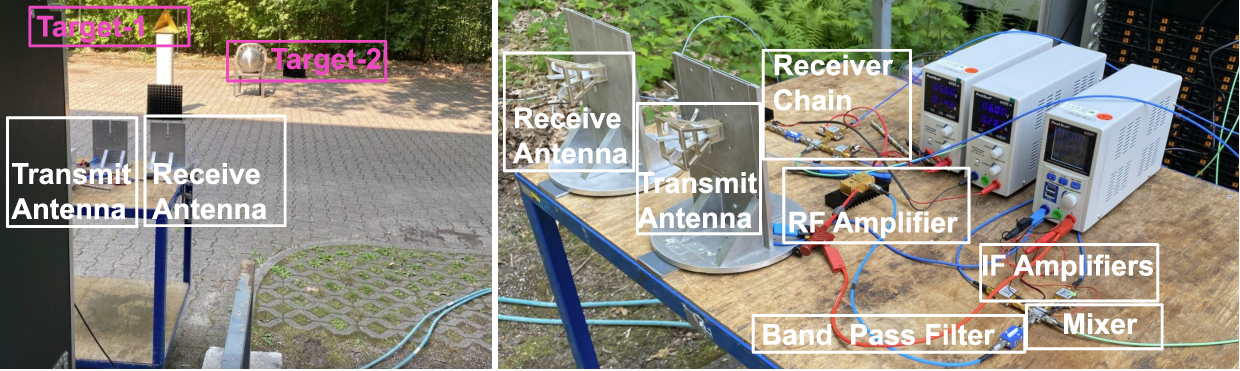}
\caption{Measurement environment used for algorithm validation. The same receive antenna is used for legitimate receiver and Eve processing.}
\label{fig:meas_setup}
\vspace{-0.2cm}
\end{figure}

\label{sec:results}
\begin{figure*}[t]
   \centering
        
        \psfrag{-50}[c][c]{\scriptsize -$50$}
        \psfrag{-45}[c][c]{\scriptsize -$45$}
        \psfrag{-40}[c][c]{\scriptsize -$40$}
        \psfrag{-35}[c][c]{\scriptsize -$35$}
        \psfrag{-30}[c][c]{\scriptsize -$30$}
        \psfrag{-25}[c][c]{\scriptsize -$25$}
        \psfrag{-20}[c][c]{\scriptsize -$20$}
        \psfrag{-15}[c][c]{\scriptsize -$15$}
        \psfrag{-10}[c][c]{\scriptsize -$10$}
        \psfrag{-5}[c][c]{\scriptsize -$5$}
        \psfrag{0}[c][c]{\scriptsize $0$}
        \psfrag{5}[c][c]{\scriptsize $5$}
        \psfrag{10}[c][c]{\scriptsize $10$}
       \psfrag{15}[c][c]{\scriptsize $15$}
        \psfrag{20}[c][c]{\scriptsize $20$}
        \psfrag{25}[c][c]{\scriptsize $25$}
        \psfrag{30}[c][c]{\scriptsize $30$}      
        \psfrag{1}[c][c]{\scriptsize $1$}
        \psfrag{2}[c][c]{\scriptsize $2$}
        \psfrag{3}[c][c]{\scriptsize $3$}
        \psfrag{4}[c][c]{\scriptsize $4$}
        \psfrag{6}[c][c]{\scriptsize $6$}
        \psfrag{7}[c][c]{\scriptsize $7$}
        \psfrag{8}[c][c]{\scriptsize $8$}
        \psfrag{XXX1}[c][c]{\scriptsize Velocity (m/s)}
        \psfrag{YYY1}[c][c]{\scriptsize Range (m)}
        \psfrag{XXX2}[c][c]{\scriptsize Velocity (m/s)}
        \psfrag{YYY2}[c][c]{\scriptsize Range (m)}
        \psfrag{XXX3}[c][c]{\scriptsize Velocity (m/s)}
        \psfrag{YYY3}[c][c]{\scriptsize Range (m)}
        \psfrag{XXX4}[c][c]{\scriptsize Velocity (m/s)}
        \psfrag{YYY4}[c][c]{\scriptsize Range (m)}
        \psfrag{ZZZ4}[c][c]{\scriptsize Normalized Power (dB)}  
        \psfrag{(a)}[c][c]{\scriptsize (a)}
        \psfrag{(b)}[c][c]{\scriptsize (b)}
        \psfrag{(c)}[c][c]{\scriptsize (c)}
        \psfrag{(d)}[c][c]{\scriptsize (d)}
        \includegraphics[width=0.9\linewidth]{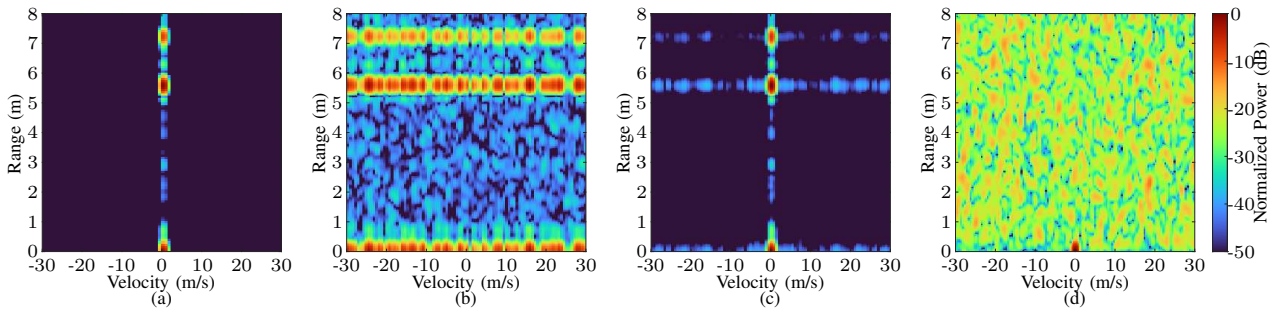}
    \caption{Radar images at (a) legitimate receiver, (b) blind Eve when \ac{OFDM}-symbol-wise phase rotations are applied, (c) smart Eve when \ac{OFDM}-symbol-wise phase rotations are applied, and (d) smart Eve when subcarrier-wise phase rotation is applied.}
    \label{fig:meas_rx_ev}
    \vspace{-0.2cm}
\end{figure*}

In this section, measurement results are presented to validate the proposed artificial-impairment-based \ac{LPI} \ac{OFDM} waveform design. Measurements were conducted at a carrier frequency of \SI{27}{\giga\hertz} with a bandwidth of \SI{1}{\giga\hertz}. The frame comprises ${M=1024}$ \ac{OFDM} symbols and ${N=6756}$ subcarriers, modulated using QPSK. A cyclic prefix of $N_\text{cp}=3378$ is appended, and pilots are inserted with frequency and time spacings of $\Delta N_\text{pil}=2$ and $\Delta M_\text{pil}=2$, respectively.

The measurements were conducted using the testbed described in \cite{nuss2024testbed}.
The measurement environment comprised two targets at different ranges, realized using a corner reflector and a dielectric sphere. The measurement setup is illustrated in Fig.~\ref{fig:meas_setup}.
To isolate and observe the \ac{LPI} strategy effectiveness, both the legitimate receiver and Eve were positioned next to the transmitter during measurements. In this configuration, spillover between transmit and receive antennas models the \ac{LoS} component shown in Fig.~\ref{fig:system}, maintaining consistency with the passive bistatic receiver model.

Fig.~\ref{fig:meas_rx_ev}a displays the measured radar image at the legitimate receiver employing the proposed \ac{LPI} waveform with compensation. Both targets are clearly resolved.
Fig.~\ref{fig:meas_rx_ev}b shows the radar image at the eavesdropper when the \ac{OFDM}-symbol-wise phase rotation and \ac{OFDM}-symbol-wise Doppler shifts are adopted, and Eve has no compensation strategy. As seen in Fig.~\ref{fig:meas_rx_ev}b, severe distortions along the velocity axis and an elevated noise floor are observed due to uncompensated phase and Doppler shifts. However, a smart-Eve can calculate the phase change through \ac{OFDM} symbols for the \ac{LoS} signal, by tracking phase changes for zero range after the range \ac{IDFT} and therefore estimating the artificial phase shifts. When this compensation strategy is adopted, the radar image shown in Fig.~\ref{fig:meas_rx_ev}c is obtained by smart-Eve. Hence, Eve can compensate the strong sidelobes in the radar image up to some extent when the \ac{OFDM}-symbol-wise phase rotation strategy was adopted at the transmitter. As shown in Fig.~\ref{fig:meas_rx_ev}c, there are still visible strong sidelobes for targets due to phase estimation errors and uncompensated artificial Doppler shifts. 
When phase rotations are applied subcarrier-wise on $\mathbf{X}$, Eve's strategy of estimating a single symbol-wise phase rotation fails, as the artificial phase shifts vary independently across the subcarriers. Even if Eve attempts the same phase-tracking strategy used in the symbol-wise case to correct phase discontinuities across OFDM symbols, the resulting radar image remains severely degraded, as shown in Fig.~\ref{fig:meas_rx_ev}d.

To compare the sensing capabilities across the four scenarios, we employ image \ac{SINR}, \ac{PSLR}, and \ac{ISLR} as performance metrics. For clarity, \ac{PSLR} is defined as the ratio of the maximum sidelobe power to the peak main lobe power, expressed in dB. 
Image \ac{SINR} is calculated in the radar image as the ratio of the peak target power to the average noise floor power in radar image. To ensure that the metric captures the degradation caused by the LPI waveform rather than environmental clutter, a mask around the zero-velocity bins ($\approx \pm \qty{5}{\meter\per\second}$) is applied to exclude clutter effects for the interference and noise level estimation. Moreover, a Chebyshev window with \qty{100}{dB} sidelobe suppression was applied during the processing to suppress sidelobes and ensure that the image \ac{SINR} degradation is attributable to the LPI waveform.
\ac{PSLR} and \ac{ISLR} are calculated along the velocity axis of the strongest target (trihedral reflector), which quantify the spreading of target energy into adjacent Doppler bins.

Table~\ref{tab:res} summarizes the measured performance metrics (SINR, PSLR, ISLR) across all scenarios shown in Fig.~\ref{fig:meas_rx_ev}. 
The numerical results confirm the observations in Fig.~\ref{fig:meas_rx_ev} and highlight the vulnerability of the baseline approach. Specifically, in the OFDM-symbol-wise impairment scenario, smart Eve achieves a viable \ac{SINR} of \qty{20.71}{dB} and a \ac{PSLR} of \qty{-14.28}{dB}, which corresponds to the clearly refocused target seen in Fig.~\ref{fig:meas_rx_ev}c. In contrast, under the proposed subcarrier-wise rotations, the smart Eve's image \ac{SINR} is suppressed to \qty{5.20}{dB}, and \ac{PSLR} degrades to \qty{-0.88}{dB}, confirms that the target energy remains indistinguishable from the noise floor, resulting in the highly distorted radar image in~Fig.~\ref{fig:meas_rx_ev}d.

\begin{table}[t]
\vspace{-0.2cm}
    \renewcommand{\arraystretch}{1.5}
    \setlength{\arrayrulewidth}{.1mm}
    \setlength{\tabcolsep}{4pt}		
    \centering
\captionsetup{width=43pc,justification=centering,labelsep=newline}
    \caption{\textsc{Performance results}}
    \label{tab:OFDM_par}
    \begin{tabular}{|c|c|c|c|}
        \hhline{|====|}
        & \textbf{SINR (dB)} & \textbf{PSLR (dB)} & \textbf{ISLR (dB)} \\\hline
        Receiver Processing & 74.81 & -66.87 & -47.91 \\\hline
        Blind Eve Processing & 2.44 & -0.90 & 14.59 \\\hline
        Smart Eve (\ac{OFDM}-symbol-wise) & 20.71 & -14.28 & 3.06 \\\hline
        Smart Eve (subcarrier-wise) & 5.20 & -0.88 & 12.12 \\\hline
        \hhline{|====|}
    \end{tabular}
    \label{tab:res}
\vspace{-0.2cm}
\end{table}


\section{Conclusion}

In this paper, an \ac{LPI} waveform design is proposed for \ac{OFDM}-based \ac{ISAC} systems using artificial phase and Doppler shifts. The approach preserves legitimate communication and sensing performance while significantly degrading eavesdropper sensing and communication intercept capability. The compensation algorithm ensures perfect recovery at the intended receiver. Measurements confirm an increased range–Doppler degradation at the eavesdropper, validating the \ac{LPI} effectiveness.

\section*{Acknowledgment}
The authors acknowledge support from the German Federal Ministry of Research, Technology and Space under grant 16KISK010.



\bibliographystyle{IEEEtran}
\bibliography{references}


\end{document}

%% file: EuMW_modify_IEEEtran_18b_CTAN_v5.tex
\makeatletter

\def\@maketitle{\newpage
\bgroup\par\addvspace{0.5\baselineskip}\centering%
\ifCLASSOPTIONtechnote
   {\bfseries\large\@IEEEcompsoconly{\sffamily}\@title\par}\vskip 1.3em{\lineskip .5em\@IEEEcompsoconly{\sffamily}\@author
   \@IEEEspecialpapernotice\par{\@IEEEcompsoconly{\vskip 1.5em\relax
   \@IEEEtitleabstractindextextbox{\@IEEEtitleabstractindextext}\par
   \hfill\@IEEEcompsocdiamondline\hfill\hbox{}\par}}}\relax
\else
   \vskip0.2em{\EuMWtitlesize\ifCLASSOPTIONtransmag\bfseries\LARGE\fi\@IEEEcompsoconly{\sffamily}\@IEEEcompsocconfonly{\normalfont\normalsize\vskip 2\@IEEEnormalsizeunitybaselineskip
   \bfseries\Large}\@title\par}\vskip1.0em\par
   \ifCLASSOPTIONconference%
      {\@IEEEspecialpapernotice\mbox{}\vskip\@IEEEauthorblockconfadjspace%
       \mbox{}\hfill\begin{@IEEEauthorhalign}\@author\end{@IEEEauthorhalign}\hfill\mbox{}\par}\relax
   \else
      \ifCLASSOPTIONpeerreviewca
         {\@IEEEcompsoconly{\sffamily}\@IEEEspecialpapernotice\mbox{}\vskip\@IEEEauthorblockconfadjspace%
          \mbox{}\hfill\begin{@IEEEauthorhalign}\@author\end{@IEEEauthorhalign}\hfill\mbox{}\par
          {\@IEEEcompsoconly{\vskip 1.5em\relax
           \@IEEEtitleabstractindextextbox{\@IEEEtitleabstractindextext}\par\hfill
           \@IEEEcompsocdiamondline\hfill\hbox{}\par}}}\relax
      \else
         \ifCLASSOPTIONtransmag
           {\@IEEEspecialpapernotice\mbox{}\vskip\@IEEEauthorblockconfadjspace%
            \mbox{}\hfill\begin{@IEEEauthorhalign}\@author\end{@IEEEauthorhalign}\hfill\mbox{}\par
           {\vspace{0.5\baselineskip}\relax\@IEEEtitleabstractindextextbox{\@IEEEtitleabstractindextext}\vspace{-1\baselineskip}\par}}\relax
         \else
           {\lineskip.5em\@IEEEcompsoconly{\sffamily}\sublargesize\@author\@IEEEspecialpapernotice\par
           {\@IEEEcompsoconly{\vskip 1.5em\relax
            \@IEEEtitleabstractindextextbox{\@IEEEtitleabstractindextext}\par\hfill
            \@IEEEcompsocdiamondline\hfill\hbox{}\par}}}\relax
         \fi
      \fi
   \fi
\fi\par\addvspace{0.0\baselineskip}\egroup}

\def\EuMWtitlesize{\@setfontsize{\EuMWtitlesize}{24}{24pt}}
\def\EuMWauthorsize{\@setfontsize{\EuMWauthorsize}{11}{11pt}}
\def\EuMWaffilsize{\@setfontsize{\EuMWaffilsize}{10}{10pt}}
\def\EuMWcaptionsize{\@setfontsize{\EuMWcaptionsize}{9}{10pt}}
\def\EuMWbibsize{\@setfontsize{\EuMWbibsize}{8}{10pt}}

\def\@IEEEauthorblockNstyle{\EuMWauthorsize\@IEEEcompsocnotconfonly{\sffamily}\@IEEEcompsocconfonly{\large}}
\def\@IEEEauthorblockAstyle{\EuMWaffilsize\@IEEEcompsocnotconfonly{\sffamily}\@IEEEcompsocconfonly{\itshape}\@IEEEcompsocconfonly{\large}}
\def\@IEEEauthordefaulttextstyle{\EuMWauthorsize\@IEEEcompsocnotconfonly{\sffamily}\sublargesize}

\def\thebibliography#1{\section*{\refname}%
    \addcontentsline{toc}{section}{\refname}%
    \EuMWbibsize\@IEEEcompsocconfonly{\small}\vskip 0.3\baselineskip plus 0.1\baselineskip minus 0.1\baselineskip
    \list{\@biblabel{\@arabic\c@enumiv}}%
    {\settowidth\labelwidth{\@biblabel{#1}}%
    \leftmargin\labelwidth
    \advance\leftmargin\labelsep\relax
    \itemsep \IEEEbibitemsep\relax
    \usecounter{enumiv}%
    \let\p@enumiv\@empty
    \renewcommand\theenumiv{\@arabic\c@enumiv}}%
    \let\@IEEElatexbibitem\bibitem%
    \def\bibitem{\@IEEEbibitemprefix\@IEEElatexbibitem}%
\def\newblock{\hskip .11em plus .33em minus .07em}%
\ifCLASSOPTIONtechnote\sloppy\clubpenalty4000\widowpenalty4000\interlinepenalty100%
\else\sloppy\clubpenalty4000\widowpenalty4000\interlinepenalty500\fi%
    \sfcode`\.=1000\relax}

\long\def\@makecaption#1#2{%
\ifx\@captype\@IEEEtablestring%
\par\@IEEEtabletopskipstrut
\else
\@IEEEfigurecaptionsepspace
\fi
\setbox\@tempboxa\hbox{\normalfont\footnotesize {#1.}\nobreakspace\nobreakspace #2}%
\ifdim \wd\@tempboxa >\hsize%
\setbox\@tempboxa\hbox{\normalfont\footnotesize {#1.}\nobreakspace\nobreakspace}%
\parbox[t]{\hsize}{\normalfont\footnotesize\noindent\unhbox\@tempboxa#2}%
\else
\ifCLASSOPTIONconference \hbox to\hsize{\normalfont\footnotesize\hfil\box\@tempboxa\hfil}%
\else \hbox to\hsize{\normalfont\footnotesize\box\@tempboxa\hfil}%
\fi\fi
\ifx\@captype\@IEEEtablestring%
\@IEEEtablecaptionsepspace
\else
\fi}

\newlength\tablecaptiontotableskip
\newlength\figuretocaptionskip
\def\@IEEEfigurecaptionsepspace{\vskip\figuretocaptionskip\relax}%
\def\@IEEEtablecaptionsepspace{\vskip\tablecaptiontotableskip\relax}%

\def\abstract{\normalfont%
\@IEEEabskeysecsize\bfseries\textit{\abstractname}\,\bfseries\textit{---}\,%
\@IEEEgobbleleadPARNLSP}%

\def\IEEEkeywords{\normalfont%
\@IEEEabskeysecsize\bfseries\textit{\IEEEkeywordsname}\,\bfseries\textit{---}\,%
\@IEEEgobbleleadPARNLSP}%
\def\endIEEEkeywords{\relax\vspace{0.67ex}%
\par\if@twocolumn\else\endquotation\fi%
\normalsize\normalfont}%

\def\@IEEEauthorblockNtopspace{0ex}
\def\@IEEEauthorblockAtopspace{1mm}
\def\tablename{Table}
\def\thetable{\arabic{table}}
\def\IEEEkeywordsname{Keywords}
\def\subsubsection{\@startsection{subsubsection}{3}{\z@}{1.5ex plus 1.5ex minus 0.5ex}%
{0.7ex plus .5ex minus 0ex}{\normalfont\normalsize\itshape}}%
\newlength{\CPheadmatchindent}%
\def\@seccntformat#1{\hbox to\CPheadmatchindent{\csname the#1dis\endcsname}\hskip 0.1em \relax}
\IEEEilabelindentA \parindent
\IEEEilabelindent \IEEEilabelindentA
\IEEEelabelindent \parindent
\IEEEdlabelindent \parindent
\IEEElabelindent \parindent
\makeatother